\documentclass[12pt,preprint]{aastex}
\usepackage{amsmath,psfig}

\shorttitle{Assessment of Fluorescence Data}
\shortauthors{Gorczyca et al.}

\begin{document}

\title{Assessment of the Fluorescence and Auger Data Base\\ used in
Plasma Modeling }

\author{T. W. Gorczyca, C. N. Kodituwakku, K. T. Korista, and O. Zatsarinny}
\affil{Department of Physics, Western Michigan University,
Kalamazoo, MI 49008, USA}
\email{gorczyca@wmich.edu}
\author{N. R. Badnell}
\affil{Department of Physics,
University of Strathclyde, Glasgow, G4 0NG, United Kingdom}
\author{E. Behar}
\affil{Physics Department, Technion, Haifa 32000, Israel}
\author{M. H. Chen}
\affil{Lawrence Livermore National Laboratory, Livermore, CA 94550, USA}
\and
\author{D. W. Savin}
\affil{Columbia Astrophysics Laboratory, Columbia University, New York, NY
10027, USA}

\begin{abstract}
We have investigated the accuracy of the $1s$-vacancy fluorescence data base
of \cite{k&m} resulting from the initial atomic physics calculations and the subsequent 
scaling along isoelectronic sequences.  In particular, we have
focused on the relatively simple Be-like and F-like $1s$-vacancy sequences. 
We find that the earlier atomic physics calculations for the oscillator 
strengths and autoionization rates of {\em singly-charged} \ion{B}{2}
and \ion{Ne}{2}
are in sufficient agreement with our present calculations.
However, the substantial charge dependence of these quantities along
each isoelectronic sequence, the incorrect configuration averaging 
used for \ion{B}{2},  and the neglect of 
spin-orbit effects (which become important at high-$Z$) all cast doubt on the reliability of 
the \cite{k&m} data for application to plasma modeling.

\end{abstract}
\keywords{atomic data -- atomic processes  -- line: formation -- X-rays:
general}

\section{Introduction}

In collisionally ionized or X-ray photoionized plasmas, high-energy electrons or photons lead to the
production of $1s$-vacancy ionic states which then decay
via sequential emission of single or multiple electrons and/or photons.
The exact strengths of these competing processes determine fundamentally
important quantities of the plasma 
such as the ionization balance and the observed spectra
of emitted and/or absorbed photons. Hence, interpreting the properties of these plasmas
requires accurate atomic physics calculations for the various
autoionization and radiative rates. Here we are interested in assessing
the accuracy of the available data base that provides such computed
(or inferred) Auger rates and fluorescence yields to the astrophysics  community. The accuracy of these atomic data are crucial to the 
interpretation of the spectra of photoionized plasmas such as are found in X-ray binaries and active galactic nuclei.  These data are also important 
for supernova remnants (SNRs) under conditions of non-equilibrium ionization
(NEI).

Two of the more widely used spectral codes for modeling photoionized plasmas 
are CLOUDY \citep{cloudy} and XSTAR \citep{kallman}.  A commonly used
code for modeling NEI in SNRs is that of \cite{bork}.
These all in turn rely on the table
of electron and photon emission probabilities compiled by \cite{k&m}.  This 
comprehensive data base considers the sequential multiple
electron and/or photon ejections for all stages of all $1s$-vacancy ions
in the periodic table up through zinc. In order to produce such a massive
array of numbers, however, certain approximations, questionable 
from a purely theoretical atomic physics standpoint,
were invoked. First, the only rigorously computed atomic rates were taken
from the early works of
\cite{mcguire69,mcguire70,mcguire71,mcguire72} for {\em singly-charged ions}, which furthermore neglected configuration interaction (CI) and spin-orbit effects.
Due to the limited computational resources available at the time,
and the approximations thus needed to perform such calculations,
even these cannot be considered as reliable as 
those that can be carried out with today's state-of-the-art 
capabilities. Second, these singly-ionized results were then 
scaled along entire isoelectronic
sequences, assuming constant autoionization rates and 
oscillator strengths;  this approximation is least valid for near-neutrals.  

A third approximation used by \cite{k&m} is that the electron and photon emission yields were computed using radiative and autoionization rates
that were configuration averaged over possible terms,
and the fluorescence yield was then given as a ratio of the averaged radiative
rate to the sum of the averaged radiative and averaged Auger rate.  For modeling purposes, however, 
this is incorrect; the actual
required value is the average of the term-specific yields - an average 
of ratios rather than a ratio of averages.
In other words,
the relative probability 
of producing each specific inner-shell-vacancy term, and its
subsequent term-specific decay,  needs to be considered,
and this was not done correctly for the data compiled by \cite{k&m}
(see also \cite{chen2} for a further discussion of this importance).

In this paper, we investigate the validity 
of the above three
 approximations in order to assess the accuracy of the resultant 
data base of \cite{k&m}.
To this end, we first study the simplest $1s$-vacancy 
system that can radiate via a $2p\rightarrow 1s$
dipole-allowed transition.  This is the removal of a $1s$ electron
from the ground-state 
B-like sequence, or rather the $1s2s^22p$ Be-like inner-shell excited sequence, 
which is investigated in the next section, and which is further 
simplified by the fact that only one electron, or one photon, can be emitted.
We follow in Section~\ref{sec:F} 
with a study of the simplest closed-(outer)shell case of F-like ions,
corresponding to $1s$ vacancies from the Ne-like sequence.  
A summary of our findings and concluding remarks 
are then given in Section~\ref{sec:Conclusion}.

\section{Case Study of the Be-Like Fluorescence Yields}
\label{sec:Be}
Inner-shell $1s$ vacancy of a Be-like ion, whether by photoionization or
electron-impact ionization of B-like ions (or by photoexcitation 
or electron-impact excitation of Be-like 
ions), results in either the $1s2s^22p(^{1}P)$ state or the $1s2s^22p(^{3}P)$ state.
From an independent particle perspective, in LS coupling, the following competing decay processes can then occur:
\begin{eqnarray}
 1s2s^22p(^{1}P) & {A_r\atop \longrightarrow} & 1s^22s^2(^1S) + \omega 
 \label{eqar2}\ ,\label{eqar}\\
 1s2s^22p(^{1,3}P)  & {A_{a1}\atop \longrightarrow} & 
1s^22s(^2S)\epsilon p(^{1,3}P)\label{eqaa1}\ ,\\
 1s2s^22p(^{1,3}P)  & {A_{a2}\atop \longrightarrow} & 
1s^22p(^2P)\epsilon s(^{1,3}P)\label{eqaa2}
\ ,
\end{eqnarray} 
that is, the $1s$-vacancy state can either
fluoresce, if it is in the $^1P$ state, with a radiative rate $A_r$, or autoionize, from either state, with a total state-dependent rate
$A_a=A_{a1}+A_{a2}$, yielding free electrons denoted by $\epsilon l$.  
(If left in the $^3P$ state the ion does not fluoresce - we consider 
CI and spin-orbit effects in the next section). 
The radiative rate $A_r$ in atomic units (1 a.u. $=4.1341\times 10^{16}\ {\rm s}^{-1}$)  is related to the 
dimensionless emission oscillator strength $f$ by
\begin{eqnarray}
A_r & = & 2\omega^2\alpha^3f\ ,
\end{eqnarray} 
where $\omega$ is the emitted photon energy in a.u. 
(1 a.u. of energy $=27.211$ eV)
and $\alpha\approx 1/137$ is the fine structure constant.
Here we define the emission oscillator strength as the absolute value
of the oscillator strength from the upper $1s2s^22p(^1P)$ term $j$ to the
lower $1s^22s^2(^1S)$ term $i$:
\begin{eqnarray}
f & \equiv & \left| f_{ji} \right|\ ,
\end{eqnarray}
and can thus be related to the absorption oscillator strength $f_{ij}$ from the lower term $i$ to the upper term $j$ via
\begin{eqnarray}
g_{i}f_{ij} & = & g_{j}\left| f_{ji} \right|\ ,
\end{eqnarray}
where $g_i=1$ and $g_j=3$ are the statistical weights of the 
initial and final Be-like terms, respectively.

Oscillator strengths are more convenient quantities to use 
along isoelectronic sequences because
they exhibit certain bounds.
Since the absorption oscillator strength 
is bounded by $0\le f_{ij}\le N_i$, where $N_i=2$ is the number of
$1s$ electrons,
the emission oscillator 
strength is bounded by $0\le f\le (g_i/g_j)N_i=2/3$ (for the present cases), 
and is a well-behaved function of the nuclear charge $Z$.  In fact,
if the hydrogenic approximation is valid, i.e., if the nuclear potential
dominates over the interelectronic repulsive potential, then the 
emission oscillator
strength is {\em independent} of $Z$, and the same is true for the autoionization rate $A_a$. Such an approximation
is  valid for highly-charged ions but not 
for lower-charged species.

The fluorescence yield $\xi$, from a given 
inner-shell vacancy state, 
is a measure 
of the relative probabilities of the radiative and autoionization
 decay pathways and is defined as
\begin{eqnarray}
\xi  &\equiv  &\frac{A_r}{A_r+A_a}
 =  \frac{\omega^2}{\omega^2 + \frac{1}{2\alpha^3}\left[\frac{A_a}{f}\right]}\ \ .
\label{eqfy}
\end{eqnarray} 
Thus it only depends on the squared transition energy $\omega^2$ 
and the ratio of the autoionization rate to the emission oscillator
strength $A_a/f$.  In the hydrogenic approximation, these
scale respectively with nuclear charge as $q^4$ and $q^0$ (i.e., independent of $q$), where $q=Z-3$ is the asymptotic ionic charge
seen by the outer-most electron of the Be-like ion \citep{cowan}.
With these scaling properties,  the expected behaviors at low-$Z$ and high-$Z$
are $\xi\approx 0$ and $\xi\approx 1$ (provided $f\ne 0$), respectively.

\subsection{Initial Populations, Configuration Interaction, and Spin-Orbit Effects}
\label{sec:photoionization}
As pointed in Section~\ref{sec:Be}, both the $^1P$ and $^3P$ terms 
can be populated after $1s$ photoionization
or electron-impact ionization.  Following \citet{cowan},
and using the sudden approximation, we have determined that the 
probability of populating each term 
can be deduced by considering the
squared recoupling coefficient 
\begin{eqnarray}
\left\vert\left\langle\left[\left(1s1s\right)\left(^1S\right)\right]2p\left(^2P\right)\left\vert
\left[\left(1s2p\right)\left(^{\left\{2{\cal S}+1\right\}}P\right)\right]
1s\left(^2P\right)\right\rangle\right\vert\right.^2 & = & \left\vert\left(-1\right)^{1+{\cal S}}\left(2{\cal S}+1\right)^\frac{1}{2}
\left\{
\begin{array}{ccc} {\cal S} & \frac{1}{2} & \frac{1}{2} \\ 
                                0 & \frac{1}{2} & \frac{1}{2} 
\end{array}\right\}\right\vert^2 \nonumber \\
& = & \frac{2{\cal S}+1}{4}\ ,
\label{eqrecoup}
\end{eqnarray}
where ${\cal S}=0$ for the $^1P$ state and ${\cal S}=1$ for the $^3P$ state.
This means that the states are populated according to their statistical 
weights,  and the $^1P$ state
is populated with a probability of $1/4$.
(In general, there also should be a recoupling coefficient
involving the orbital angular momenta of the three electrons in Eq.~\ref{eqrecoup};
however, the $l=0$ values for two of the electrons' orbital momenta
reduces the coefficient to unity for the present case.)
We have  also verified this computationally by performing R-matrix 
photoionization calculations
using the Wigner-Eisenbud
R-matrix method \citep{burke,berr}.  
Using both approaches we find that in 
intermediate coupling, the states are also populated according to their
statistical weights (a similar expression to Eq.~\ref{eqrecoup} involving
the total angular momentum values $j$ for each electron can be obtained).

Considering the relative populations of the $1s2s^22p$ 
$1s$-vacancy states, the desired quantity for plasma modeling purposes 
is the configuration-average fluorescence yield.
If CI and spin-orbit effects
are neglected, this can be defined as an average over LS single-configuration (SC) terms as
\begin{eqnarray}
\xi_{LSSC}  & \equiv  &\frac{1}{4}\xi{(^1P)} + \frac{3}{4}\xi{(^3P)} \nonumber\\
&=& \frac{1}{4}\xi{(^1P)}\nonumber \\
& {\longrightarrow\atop z\rightarrow\infty} & \frac{1}{4}\ ,
\label{eqca}
\end{eqnarray} 
where fluorescence from the $^3P$ state is 
zero so that the asymptotic
behavior at large $Z$ is 
$1/4$.  

CI and spin-orbit effects modify this behavior, however.
The largest CI effect is the intrashell mixing
$c_11s2s^22p+c_21s2p^3$, where the mixing fraction $\vert c_2/c_1\vert^2$
is essentially term independent and $Z$ independent 
for nonrelativistic calculations -
it varies between 0.067 for \ion{B}{2} and 0.053 for \ion{Zn}{27}.
This mixing affects the computed emission oscillator strength $f$
and autoionization rate $A_a$ at the near neutral end of the sequence,
but changes the high-$Z$ fluorescence yield by less than 10\%.  
The more important CI effect is that the admixture of the $1s2p^3$
configuration in the $^3P$ term allows it to radiate to the $1s^22p^2(^3P)$
state.
This $c_21s2p^3(^3P)\rightarrow 1s^22p^2(^3P)$ radiative rate is about a factor of 20 smaller than the
$1s2s^22p(^1P)\rightarrow 1s^22s^2(^1S)$ rate, so it only increases
the fluorescence yield by a few percent at low $Z$.  As $Z$ increases,
however,
eventually even this reduced radiative rate dominates the autoionization rate,
giving
\begin{eqnarray}
\xi_{LSCI}  & \equiv  &\frac{1}{4}\xi{(^1P)} + \frac{3}{4}\xi{(^3P)} \nonumber\\
& {\longrightarrow\atop z\rightarrow\infty} & 1\ .
\label{eqca2}
\end{eqnarray} 

The spin-orbit interaction also affects the computed fluorescence yield,
primarily by
mixing  the $^1P_1$ and $^3P_1$ levels.
The mixing fraction, while only about $6.3\times 10^{-6}$ at $Z=5$,
has a $Z^4$ dependence, and eventually becomes quite significant,
reaching $0.117$ at $Z=30$.  As a result, the ``$^3P_1$'' level   
(this is now just a label used to indicate the dominant term of a level)
has an increased fluorescence yield, and we get that the 
intermediate coupling (IC), configuration-averaged fluorescence yield,
including CI, behaves as
\begin{eqnarray}
\xi_{ICCI}  & \equiv  &\frac{3}{12}\xi{(^1P_1)} + \frac{1}{12}\xi{(^3P_0)} + \frac{3}{12}\xi{(^3P_1)}+\frac{5}{12}\xi{(^3P_2)} \nonumber\\
& \ge        & \xi_{LSCI}\ .
\label{eqca3}
\end{eqnarray}
Thus we see that CI and the spin-orbit interaction each cause 
an increase in the computed fluorescence yield as $Z$ is increased.

\subsection{Earlier Be-like Fluorescence Data}
\label{sec:km}
The approach of \cite{k&m} for this particular Be-like series was to
neglect spin-orbit and CI effects,
 and to assume that the hydrogenic approximation
is valid throughout the series.  Furthermore,
they used configuration-averaged values for the \ion{B}{2} 
autoionization rate and emission oscillator strength, which were computed by
 \cite{mcguire69}, and the 
experimental values of $\omega$ from \cite{lotz67,lotz68}, to 
obtain the ratio $A_r/(A_r+A_a)$ required for determining $\xi$ using Eq.~\ref{eqfy}.  This is not the
same as the desired configuration-averaged
fluorescence yield $\xi_{LSSC}$ in Eq.~\ref{eqca} -
the ratio of the averages does not equal the average of the ratios:
\begin{eqnarray}
{\sum_{{\cal S}=0,1}({2{\cal S}+1\over 4})A_r(^{2{\cal S}+1}P)\over 
\sum_{{\cal S}=0,1}({2{\cal S}+1\over 4})\left[A_r(^{2{\cal S}+1}P)+A_a(^{2{\cal S}+1}P)\right]}
& \ne & 
\sum_{{\cal S}=0,1}\left({2{\cal S}+1\over 4}\right)\left[{A_r(^{2{\cal S}+1}P)\over A_r(^{2{\cal S}+1}P)+A_a(^{2{\cal S}+1}P)}\right]
\ .
\label{eqne}
\end{eqnarray}
   
We first address the accuracy of the
computed autoionization rates and emission oscillator strengths 
in the next subsection,  and then address 
the validity of the hydrogenic approximation
in the following subsection.  Fluorescence yields are
presented in the last subsection, where the incorrect averaging and neglect 
of CI and spin-orbit effects by \cite{k&m} are addressed.

\subsection{Atomic Calculations for \ion{B}{2}}
\label{sec:mcguire}

In order to calculate the transition matrix elements appearing in the
expressions for the radiative and autoionization rates \citep{cowan}, it is
first necessary to produce atomic wave functions.  \cite{mcguire69}
used the Herman-Skillman approximation in determining the
(single-configuration) wave functions, whereby {\em all} electrons (i.e., 
the $1s$, $2s$, $2p$, and continuum ones) are eigenfunctions of a common
central potential;  as stated by \cite{mcguire69},
this ``neglect(s) ... exchange and correlation effects.''  
Furthermore, this potential $rV(r)$
is approximated by ``a series of straight lines'' in order to yield
piece-by-piece analytic Whitakker functions. 
Here we are concerned with the validity of these approximations,
given that more rigorous calculations can be easily performed using 
today's state-of-the-art technologies.

For the present study, we use the 
program AUTOSTRUCTURE
\citep{auto}, which generates Slater-type $1s$, $2s$, $2p$, and distorted-wave continuum orbitals.  
In order to compare with the results of \cite{mcguire69} for \ion{B}{2},
and with \cite{k&m} as we scale from $Z=5$ to $Z=30$,
we first performed single-configuration LS calculations.
For the more rigorous calculations that we compare to other theoretical
results and that we recommend as the definitive data, 
we also included CI -
$1s2s^22p\ +\ 1s2p^3$ for the inner-shell vacancy state and 
$1s^22s^2\ +\ 1s^22p^2$ for the final radiative decay state -
and spin-orbit effects.
The two accessible continua were described as $1s^22s\epsilon p$ and 
$1s^22p\epsilon s$, where $\epsilon l$ denotes a continuum distorted wave.

Given atomic wave functions, \cite{mcguire69} computed
the {\em configuration average} (CA) radiative and 
partial autoionization rates in Eqs.~1-3.
The emission oscillator strength given is thus
\begin{eqnarray}
f(CA) & = & \frac{1}{4}f(^1P) + \frac{3}{4}f(^3P) \nonumber\\
 & = & \frac{1}{4}f(^1P) \nonumber \\
 & = & 0.0377\label{eqfca} \ ,
 \end{eqnarray}
whereas for the total autoionization rate, 
the CA rates for the processes in Eqs.~2-3 were used, that is,
\begin{eqnarray}
A_a(CA) & = & A_{a1}(CA) + A_{a2}(CA)\nonumber\\
 & = & 2.37\times 10^{-3}\ {\rm a.u.}\ ,
\end{eqnarray}
where
\begin{eqnarray}
A_{a1}(CA) & = & \frac{1}{4}A_{a1}(^1P) + \frac{3}{4}A_{a1}(^3P)\nonumber\\
 & = & \frac{1}{4}\left\{
 2\pi\left[R_0(1s,\epsilon p,2s,2p)-\frac{2}{3}R_1(1s,\epsilon p,2p,2s)\right]^2
 \right\}\nonumber\\
& &  +
 \frac{3}{4}\left\{
 2\pi\left[R_0(1s,\epsilon p,2s,2p)\right]^2
 \right\} \ .
\label{eqaa1new}
\end{eqnarray}
and
\begin{eqnarray}
A_{a2}(CA) & = & \frac{1}{4}A_{a2}(^1P)+\frac{3}{4}A_{a2}(^3P) 
\nonumber\\ 
& = & 2\pi\left[R_0(1s,\epsilon s,2s,2s)\right]^2\ ,
\label{eqaa2new}
\end{eqnarray}
since our calculations indicate that $A_{a2}(^1P)=A_{a2}(^3P)$. 
Here $R_\lambda(n_1l_1,n_2l_2,n_3l_3,n_4l_4)$ is a 
Slater integral of multipole $\lambda$ \citep{cowan},
and $\epsilon l$ represents the outgoing $l$-wave continuum electron
orbital. 
(The expressions 
in Eqs.~\ref{eqaa1new} and \ref{eqaa2new} are equivalent to those in Eq. 6 of \cite{mcguire69} for inequivalent electrons and
single-$p$ orbital occupation, considering the different
continuum normalization used by \cite{mcguire67}).
Note that the partial rate $A_{a1}(^1P)$ in Eq.~\ref{eqaa1new} is greatly suppressed relative
to the $A_{a1}(^3P)$ rate due to a near cancellation of monopole and dipole
Slater integrals.
(Indeed, it was due to this near cancellation of Slater integrals
that \cite{caldwell}
explained why the inner-shell photoexcited $1s2s^22p(^1P)$ 
resonance in \ion{Be}{1}
preferentially decayed - by two orders of magnitude - to the $1s^22p(^2P)+e^-$
channel, compared to the $1s^22s(^2S)+e^-$
channel.)
Thus the configuration average partial rate $A_{a1}(CA)$ will be larger than
the partial rate $A_{a1}(^1P)$, and hence the configuration average total 
rate $A_a(CA)$ will be larger than $A_{a}(^1P)$.  

Since we are interested in computing $\xi(^1P)$, which requires $A_a(^1P)$
and $f(^1P)$, we have converted the reported values from \cite{mcguire69}
to the $^1P$ values (the Slater integrals were also given in that
work).  We get the following values
\begin{eqnarray}
f(^1P,{\rm McGuire}) & = & 0.1508\\
A_a(^1P,{\rm McGuire}) & = & 1.692\times 10^{-3}\ {\rm a.u.}\ ,
\end{eqnarray}
which compare fairly well with our results obtained using AUTOSTRUCTURE:
\begin{eqnarray}
f(^1P,{\rm present}) & = & 0.1519\\
A_a(^1P,{\rm present}) & = & 1.045\times 10^{-3}\ {\rm a.u.}
\end{eqnarray}

In summary, we find that the earlier results for \ion{B}{2} of \cite{mcguire69} 
are consistent with ours.  
However, for the astrophysical plasma modeling purposes 
we have in mind, one really requires the configuration average fluorescence
yield, not the ratio of the averaged radiative and total rates used by \cite{k&m}
\begin{eqnarray}
\xi(K\&M) & = & \frac{A_r(CA)}{A_r(CA)+A_a(CA)}\nonumber \\   
 & {\longrightarrow\atop z\rightarrow\infty}  & 1\ ,   
\label{eqxikm}
\end{eqnarray}
due to $A_r$ and $A_a$ scaling as $q^4$ and $q^0$, respectively, in the
hydrogenic approximation.
Equation~\ref{eqxikm} differs from
the correct  $\xi_{LSSC}$ given in Eq. \ref{eqca}.  
First, we have $A_a(CA) > A_a(^1P)$ due to the near cancellation
in the $^1P$ $2s2p\rightarrow 1s\epsilon p$ partial autoionization rate,
so at low $Z$, where $A_r\ll A_a$, we have $\xi(K\&M)< \xi_{LSSC}$. 
Second,  when CI and spin-orbit effects are ignored,
as they were in \cite{k&m},
the fluorescence yields differ asymptotically by a factor of 4,
\begin{eqnarray}
\lim_{Z\rightarrow\infty}\frac{\xi(K\&M)}{\xi_{LSSC}} & = & 4\ ,
\end{eqnarray}
as can be seen by comparing Eqs.~\ref{eqca} and \ref{eqxikm}.
Of course, CI needs to be included for all $Z$, whereas spin-orbit mixing needs to be
included at higher $Z$, 
and both $\xi(K\&M)\rightarrow 1$ and $\xi_{ICCI}\rightarrow 1$ 
as $Z\rightarrow \infty$.  However, 
in the intermediate $Z$ range, it can be shown that
the \cite{k&m} results are still larger than the ICCI results. 

\subsection{Validity of the Hydrogenic Approximation}
\label{sec:scaling}
In order to assess the validity of scaling the \ion{B}{2}
results along the isoelectronic series, we computed both 
the $^1P$ autoionization rate $A_a$ and emission oscillator strength $f$ 
for all Be-like ions up through zinc, first neglecting
spin-orbit effects.  In Fig.~1, it is seen that neither
of the two is independent of the nuclear charge $Z$ at the lowest stages
of ionization - the emission oscillator strength increases by about $2/3$ in going toward
the highly-ionized regime whereas the autoionization rate more than doubles.  Furthermore, by choosing the scale so that
our two quantities coincide for \ion{B}{2}, it is seen that the
important ratio $A_a/f$ appearing in Eq.~\ref{eqfy} increases by roughly 25\% 
by the time \ion{Zn}{27} is reached.  Thus the assumption of
pure hydrogenic scaling by \cite{k&m} alone introduces an 
uncertainty at the highly-charged end of this series.
Due to the stronger $Z$ dependence at the near-neutral end,
together with the greater sensitivity to the atomic basis used in this region,
we recommend that if scaling along an isoelectronic sequence is to be performed,
the better starting point would be at the highest $Z$ desired, extrapolating
the rates
to lower $Z$ members.  Of course, given the ease of determining
atomic rates with modern computing capabilities, the most reliable approach 
is to calculate the fluorescence yield directly rather than resort to
questionable scaling methods.

\subsection{Fluorescence Yield Results}
\label{sec:Results}
While the assumption of hydrogenic scaling introduces an
$\approx 25\%$ inaccuracy in $A_a/f$, the initial quantity being scaled in \cite{k&m} - the ratio of averages rather than the average of ratios -
is really not the desired quantity to be scaled in the first place.
 Together, these approximations lead to an uncertain prediction for the 
fluorescence yield.  In Fig.~2 (and Table~1),
we compare various results for $\xi$ along the Be-like
sequence, where it can be seen that our single-configuration LS results
differ greatly from those of \cite{k&m}, especially at higher $Z$;
here, especially, their results are expected to differ from the
correct single-configuration
values due to their incorrect asymptotic value given by 
Eq.~\ref{eqxikm}.
A more disturbing result was found when we tried to repeat 
their calculations, i.e., when we
used Eq.~\ref{eqfy},
with the ratio of $A_a(CA)/f(CA)$ taken from \cite{mcguire69}, and the energies
$\omega$ taken from \cite{lotz67,lotz68}.  Whereas these scaled results
exhibit a smooth monotonic increase with nuclear charge $Z$, those
of \cite{k&m} are somewhat irregular, showing unphysical dips, and do not
agree with what we tried to reproduce, given their stated method.
Either way, the results of \cite{k&m}, or our scaled ones using
the \ion{B}{2} results of \cite{mcguire69}, initially 
underestimate our results
at lower $Z$, and then overestimate our
(LSSC) results by almost a factor of 3 for the highest $Z=30$. 

To our knowledge, there have been two other calculations
for the fluorescence yields of some members of the Be-like
sequence: those of \cite{behar} using the HULLAC codes \citep{BarS98a}
and those of \cite{chen} using a multiconfiguration Dirac-Fock (MCDF) 
method.  In both cases, CI and spin-orbit effects were included.
Here we do the same, first adding the important $2s^2\rightarrow 2p^2$
CI discussed earlier to the LS calculations in order to see
that this effect increases the $Z=30$ fluorescence yield 
by about 30\%.  Then when spin-orbit effects (and other higher-order,
relativistic effects) are included in our intermediate coupling
calculation, there is a further increase in the fluorescence
yield by about 20\% more.
In comparison with the other two calculations along this series, there is 
overall good agreement with these IC results.

\section{Case Study of the F-Like Fluorescence Yields}
\label{sec:F}

We turn now to the simplest closed-(outer)shell case of a
$1s$-vacancy in F-like ions, giving the $1s2s^22p^6(^2S)$ state
which decays as 
\begin{eqnarray}
 1s2s^22p^6(^2S) & {A_r\atop \longrightarrow} & 1s^22s^22p^5(^2P) + \omega 
 \label{eqarf}\\
 & {A_a\atop \longrightarrow} & \left\{\begin{array}{l}
1s^22p^6(^1S)\epsilon s\\
1s^22s2p^5(^{1,3}P)\epsilon p\\
1s^22s^22p^4(^{3}P,\, ^1D,\, ^1S)\epsilon s, \epsilon d
\end{array}\right.
\ .
\label{eqfauto}
\end{eqnarray} 
Again, only one photon, or one electron, can be emitted, which simplifies the 
analysis considerably (when spin-orbit effects are considered,
the final ionic term in Eq.~\ref{eqarf} is fine structure split into
 the ground $1s^22s^22p^5(^2P_{3/2})$ level
and the metastable $1s^22s^22p^5(^2P_{1/2})$ level).  Since this is a closed-shell
system, the Herman-Skillman method for the important $2p$ electrons
is expected to be more accurate than for \ion{B}{2}.  Indeed, as stated by \cite{mcguire69},
``in stripping away electrons (in reducing to a closed-shell system),
... we should be increasing the applicability of the common central-field
approximation.''  Furthermore, there is only one $1s2s^22p^6$  $1s$-vacancy 
state,
rather than the two we had for the Be-like sequence, 
and no other intrashell configurations to CI mix with, 
so we do not need to consider population of non-fluorescing states by CI or spin-orbit mixing,
nor do we have to consider configuration averaging issues.  Consequently, 
a single configuration LS coupling calculation is sufficient to
determine accurate $A_a$, $f$, and $\xi$ values for the $^2P$ term.

As a result, the computed values of the autoionization
rate and emission oscillator strength given by \cite{mcguire69} agree quite well with our
values, as seen in Fig.~3 and Table~2.  However, both of these values
depend on the internuclear charge $Z$, giving a ratio $A_a/f$ 
that increases by about a factor of $1/2$ in going
from \ion{Ne}{2} to \ion{Zn}{22}.  Thus the scaled fluorescence yield
$\xi$, using Eq.~\ref{eqfy}, the ratio $A_a/f$ from \cite{mcguire69},
and $\omega$ from \cite{lotz67,lotz68}, increases relative to the actual computed value, as is seen in Fig.~4 and Table~2.  The more troublesome news in this 
figure is the actual tabulated values of \cite{k&m} - their values 
do not follow 
our attempt at reproducing those results, but rather tend to follow 
our computed values, except for certain unphysical dips.  Nevertheless, 
the results reported by \cite{k&m} for F-like ions
are not plagued by as many uncertainties as those for Be-like ions.
We also see in Fig.~4 that the HULLAC results are in good agreement
with our present ones (the results reported earlier by \citet{behar}
only considered fluorescence into the $1s^22s^22p^5(^2P_{3/2})$ level, 
which includes only 4 of all 6 magnetic sublevels of the $1s^22s^22p^5(^2P)$
configuration;
therefore, those values must 
be multiplied by about 3/2 to account for fluorescence into
the two $1s^22s^22p^5(^2P_{1/2})$ sublevels as well.  Furthermore, the earlier HULLAC result for F$^+$ 
was erroneously listed incorrectly, and here we have given 
the actual computed value that should have appeared).

\section{Summary and Conclusion}
\label{sec:Conclusion}
The inaccuracies we have discovered
in the reported results of \cite{k&m} for Be-like ions are as follows:
\begin{enumerate}
\item The computed atomic data for \ion{B}{2} 
are used in the form $A_a(CA)/f(CA)$, that is, the radiative and autoionization rates have been averaged over the $^1P$ and $^3P$ configurations, whereas the
desired quantity for plasma modeling applications
is $\xi_{ICCI}$ and is not the same thing, differing qualitatively
and quantitatively, especially
in the asymptotic high-$Z$ limit.
\item The hydrogenic scaling assumed is invalid.  The autoionization rates, 
the emission oscillator strengths, and even the ratio $A_a/f$ are not independent of nuclear 
charge $Z$.  
\item The tabulated data of \cite{k&m} do not seem to follow the results we 
obtain when we try to reproduce their stated method using Eq.~\ref{eqfy},
with $A_a(CA)/f(CA)$ from \cite{mcguire69}, and $\omega$ from \cite{lotz67,lotz68}.
\item The calculations of \cite{k&m} neglected CI and spin-orbit effects as they scaled to higher $Z$.
\end{enumerate}

For F-like ions, items 1 and 4 are not issues since there is only one 
inner-shell vacancy term.  However, points 2 and 3 still apply for the F-like sequence. For plasma modeling purposes, we recommend 
our $\xi_{ICCI}$ for the Be-like sequence and our $\xi$ for the F-like
sequence.

In conclusion, we propose that,
given the many uncertainties discovered,
the entire data base of \cite{k&m} should be reevaluated.
While we have focused on systems that can emit only one photon or one electron,
their comprehensive tabulation also includes data for ions with $n\ge 3$ shells 
occupied;
these can emit multiple electrons and/or photons through numerous
cascading channels, compounding the inaccuracies we have discovered. 

\acknowledgements

We would like to thank E. J. McGuire for careful reading of,
and extremely helpful comments on, an earlier version of this manuscript.
TWG, CNK, KTK, and OZ were supported by NASA Space Astrophysics
Research and Analysis Program grant NAG5-10448.  
EB was supported by the Yigal-Alon Fellowship and by the GIF Foundation under grant \#2028-1093.7/2001. 
The work of MHC was performed under the auspices of US Department of 
Energy by the University of California, Lawrence Livermore National 
Laboratory, Under Contract Number W-7405-ENG-48.
DWS was supported in part by NASA Space
Astrophysics Research and Analysis Program grant NAG5-5261
and NASA Solar Physics research, Analysis, and Suborbital Program grant NAG5-9581.

\vfill
\clearpage
\eject

\begin{deluxetable}{rllcccccccc}
\tabletypesize{\scriptsize}
\tablecaption{Emission oscillator strengths, autoionization rates, fluorescence yields,
and photon energies
for Be-like $1s2s^22p$ ions.
\label{table1}}
\tablewidth{0pt}
\tablehead{
\colhead{$Z$} & \colhead{$f$} & \colhead{$A_a$} & \colhead{$\xi$} & \colhead{$\xi$} & \colhead{$\omega$} & \colhead{$\xi$} & \colhead{$\xi$}& \colhead{$\xi$} & \colhead{$\xi$} & \colhead{$\xi$}
\\
 & present\tablenotemark{a} & present\tablenotemark{b} & present LSSC\tablenotemark{c} & K\&M\tablenotemark{d} & Lotz\tablenotemark{e} & Scaled\tablenotemark{f} &
 HULLAC\tablenotemark{g} & MCDF\tablenotemark{h} & present LSCI\tablenotemark{i} & present ICCI\tablenotemark{j} }
\startdata
  5 &  0.1519 &  0.1045 &  0.0014 & 0.0006 &    6.751 &   0.0006 &
  &  &  0.0011 & 0.0011\\
    &  0.1508\tablenotemark{k} &  0.1692\tablenotemark{k}  &   &    &     &    \\
  6 &  0.1712 &  0.1194 &  0.0032  &   0.0019 &   10.349 &   0.0013 &
  & 0.0024  &  0.0025 & 0.0025\\
  7 &  0.1859 &  0.1328 &  0.0061  &   0.0052 &   14.721 &   0.0027 &
  & 0.0045  &  0.0048 & 0.0048\\
  8 &  0.1972 &  0.1442 &  0.0106  &   0.0096 &   19.866 &   0.0049 &
  & 0.0079  &  0.0083 & 0.0083\\
  9 &  0.2062 &  0.1540 &  0.0168  &   0.0154 &   25.753 &   0.0081 &
  & 0.0128  &  0.0132 & 0.0133\\
 10 &  0.2134 &  0.1622 &  0.0250  &   0.0229 &   32.379 &   0.0128 &
    0.0209 & 0.0191  &  0.0199 & 0.0201\\
 11 &  0.2193 &  0.1693 &  0.0352  &   0.0352 &   39.782 &   0.0192 &
  &  &  0.0285 & 0.0287\\
 12 &  0.2243 &  0.1754 &  0.0474  &   0.0424 &   47.924 &   0.0276 &
  0.0414 & 0.0377  &  0.0390 & 0.0393\\
 13 &  0.2285 &  0.1808 &  0.0612  &   0.0484 &   56.806 &   0.0384 &
  0.0538 &  &  0.0514 & 0.0518\\
 14 &  0.2320 &  0.1854 &  0.0761  &   0.0768 &   66.465 &   0.0518 &
  0.0685 &  &  0.0653 & 0.0658\\
 15 &  0.2351 &  0.1896 &  0.0916  &   0.1102 &   76.862 &   0.0681 &
  &  &  0.0805 & 0.0812\\
 16 &  0.2378 &  0.1933 &  0.1073  &   0.1446 &   88.034 &   0.0874 &
  0.0984 &  &  0.0965 & 0.0974\\
 17 &  0.2402 &  0.1965 &  0.1225  &   0.1656 &   99.982 &   0.1100 &
  &  &  0.1129 & 0.1141\\
 18 &  0.2423 &  0.1995 &  0.1369  &   0.1671 &  112.664 &   0.1357 &
  0.1273 & 0.1237  &  0.1295 & 0.1309\\
 19 &  0.2442 &  0.2022 &  0.1502  &   0.1626 &  126.122 &   0.1644 &
  &  &  0.1458 & 0.1478\\
 20 &  0.2459 &  0.2046 &  0.1623  &   0.1984 &  140.348 &   0.1958 &
  0.1569 &   &  0.1616 & 0.1646\\
 21 &  0.2475 &  0.2068 &  0.1732  &   0.2963 &  155.342 &   0.2298 &
  &  &  0.1769 & 0.1813\\
 22 &  0.2488 &  0.2089 &  0.1828  &   0.3438 &  171.107 &   0.2658 &
  &  &  0.1916 & 0.1982\\
 23 &  0.2501 &  0.2108 &  0.1912  &   0.3838 &  187.645 &   0.3033 &
  &  &  0.2058 & 0.2154\\
 24 &  0.2513 &  0.2125 &  0.1985  &   0.4214 &  204.991 &   0.3419 &
  &  &  0.2194 & 0.2333\\
 25 &  0.2523 &  0.2141 &  0.2049  &   0.4562 &  223.108 &   0.3810 &
  &  &  0.2327 & 0.2518\\
 26 &  0.2533 &  0.2156 &  0.2105  &   0.4903 &  241.998 &   0.4200 &
  0.2394 & 0.2633  &  0.2457 & 0.2713\\
 27 &  0.2542 &  0.2169 &  0.2153  &   0.5267 &  261.659 &   0.4584 &
  &  &  0.2585 & 0.2916\\
 28 &  0.2551 &  0.2182 &  0.2194  &   0.5836 &  282.129 &   0.4960 &
  &  &  0.2712 & 0.3125\\ 
 29 &  0.2559 &  0.2194 &  0.2230  &   0.6215 &  303.333 &   0.5322 &
  &  &  0.2840 & 0.3339\\ 
 30 &  0.2566 &  0.2205 &  0.2261  &   0.6322 &  325.310 &   0.5668 &
  &  &  0.2969 & 0.3553\\ 
 \pagebreak
\enddata
\tablenotetext{a}{Present LS results for emission from the $^1P$ term (dimensionless).}
\tablenotetext{b}{Present LS results, autoionization
 from the $^1P$ term (in units of
$10^{-2}$ a.u., 1 a.u.$=4.13\times 10^{16}$ s$^{-1}$).}
\tablenotetext{c}{Present LS results using a single configuration, one fourth the
$^1P$ term fluorescence yield (dimensionless).}
\tablenotetext{d}{\citet{k&m}.}
\tablenotetext{e}{\citet{lotz67,lotz68}  (in a.u., 1 a.u.$=27.211$ eV).}
\tablenotetext{f}{Obtained using Eq.~\ref{eqfy} with $A_a(CA)/f(CA)$ for \ion{B}{2} from \citet{mcguire69} and $\omega$ from \citet{lotz67,lotz68}.}
\tablenotetext{g}{\citet{behar}, averaged over the $^1P_1$ and $^3P_{0,1,2}$
levels.}
\tablenotetext{h}{\citet{chen}, averaged over the $^1P_1$ and $^3P_{0,1,2}$
levels.}
\tablenotetext{i}{Present LS results, including configuration interaction (CI),
averaged over the $^1P$ and $^3P$ terms.}
\tablenotetext{j}{Present intermediate-coupling (IC) results, including CI,
averaged over the $^1P_1$ and $^3P_{0,1,2}$
levels.}
\tablenotetext{k}{\citet{mcguire69}.}
\end{deluxetable}

\begin{deluxetable}{rllccccc}
\tabletypesize{\scriptsize}
\tablecaption{Emission oscillator strengths, autoionization rates, fluorescence yields,
and photon energies
for F-like $1s2s^22p^6(^2S)$ ions.
\label{table2}}
\tablewidth{0pt}
\tablehead{
\colhead{$Z$} & \colhead{$f$} & \colhead{$A_a$} & \colhead{$\xi$} & \colhead{$\xi$} & \colhead{$\omega$} & \colhead{$\xi$}& \colhead{$\xi$}
\\
 & present\tablenotemark{a} & present\tablenotemark{b} & present\tablenotemark{a} & K\&M\tablenotemark{c} & Lotz\tablenotemark{d} & Scaled\tablenotemark{e} &
 HULLAC\tablenotemark{f}} 
\startdata
 10 &  0.2159 &  0.1056 & 0.0147  &   0.0182 &  31.184 & 0.0169 & 0.0215\\
 & 0.216\tablenotemark{g} & 0.0948\tablenotemark{g} & & & & &\\
 11 &  0.2286 &  0.1164 & 0.0214  &   0.0263 &  38.723 & 0.0258 & \\
 12 &  0.2406 &  0.1273 & 0.0298  &   0.0346 &  47.035 & 0.0376 & 0.0380\\
 13 &  0.2515 &  0.1377 & 0.0402  &   0.0397 &  56.081 & 0.0526 & 0.0493\\
 14 &  0.2615 &  0.1477 & 0.0528  &   0.0449 &  65.864 & 0.0712 & 0.0630\\
 15 &  0.2705 &  0.1571 & 0.0679  &   0.0634 &  76.422 & 0.0936 & \\
 16 &  0.2786 &  0.1659 & 0.0855  &   0.0875 &  87.720 & 0.1197 & 0.0983\\
 17 &  0.2859 &  0.1741 & 0.1058  &   0.1019 &  99.795 & 0.1497 & \\
 18 &  0.2926 &  0.1817 & 0.1286  &   0.1305 & 112.646 & 0.1832 & 0.1443\\
 19 &  0.2987 &  0.1888 & 0.1540  &   0.1253 & 126.276 & 0.2199 & \\
 20 &  0.3042 &  0.1954 & 0.1818  &   0.1505 & 140.682 & 0.2592 & 0.2001\\
 21 &  0.3093 &  0.2016 & 0.2118  &   0.2073 & 155.863 & 0.3004 & \\
 22 &  0.3139 &  0.2073 & 0.2437  &   0.2411 & 171.820 & 0.3429 & \\
 23 &  0.3182 &  0.2127 & 0.2771  &   0.2751 & 188.552 & 0.3860 & \\
 24 &  0.3221 &  0.2177 & 0.3116  &   0.3068 & 206.093 & 0.4289 & \\
 25 &  0.3258 &  0.2224 & 0.3469  &   0.3386 & 224.395 & 0.4710 & \\
 26 &  0.3291 &  0.2269 & 0.3825  &   0.3692 & 243.504 & 0.5118 & 0.4041\\
 27 &  0.3324 &  0.2309 & 0.4180  &   0.3942 & 263.386 & 0.5513 & \\
 28 &  0.3353 &  0.2348 & 0.4531  &   0.4438 & 284.040 & 0.5883 & \\
 29 &  0.3380 &  0.2385 & 0.4874  &   0.4734 & 305.465 & 0.6230 & \\
 30 &  0.3406 &  0.2420 & 0.5207  &   0.4758 & 327.735 & 0.6554 & \\
 \pagebreak
\enddata
\tablenotetext{a}{Present results (dimensionless).}
\tablenotetext{b}{Present results (in units of
$10^{-1}$ a.u., one a.u.$=4.13\times 10^{16}$ s$^{-1}$).}
\tablenotetext{c}{\citet{k&m}.}
\tablenotetext{d}{\citet{lotz67,lotz68} (in a.u., 1 a.u.$=27.211$ eV).}
\tablenotetext{e}{Obtained using Eq.~\ref{eqfy} with $A_a(CA)/f(CA)$ for \ion{B}{2} from \citet{mcguire69} and $\omega$ from \citet{lotz67,lotz68}.}
\tablenotetext{f}{\citet{behar}.}
\tablenotetext{g}{\citet{mcguire69}.}
\end{deluxetable}

\vfill
\clearpage
\begin{figure*}
\centerline{\psfig{file=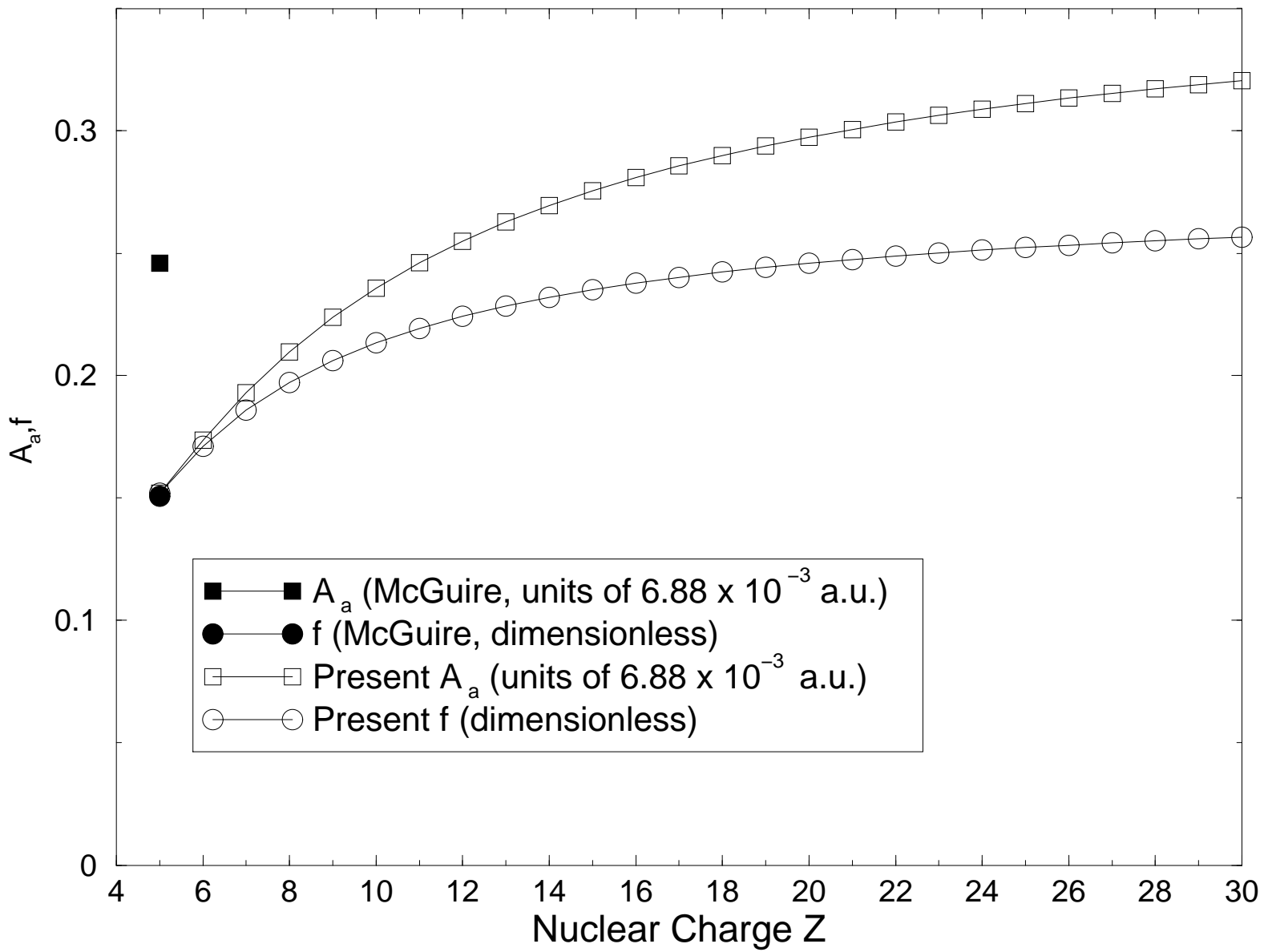,angle=-90.}}
\caption{Present LS autoionization rates
$A_a$ (in units of $6.88\times 10^{-3}$ a.u., open squares) and
emission oscillator strengths $f$ (dimensionless, open circles) 
for Be-like $1s2s^22p(^1P)$ ions
as a function
of the nuclear charge $Z$.  The autoionization rate and emission oscillator strength
from \cite{mcguire69} for 
\ion{B}{2} are shown by the solid square and circle,
respectively.}
\end{figure*}

\begin{figure*}
\centerline{\psfig{file=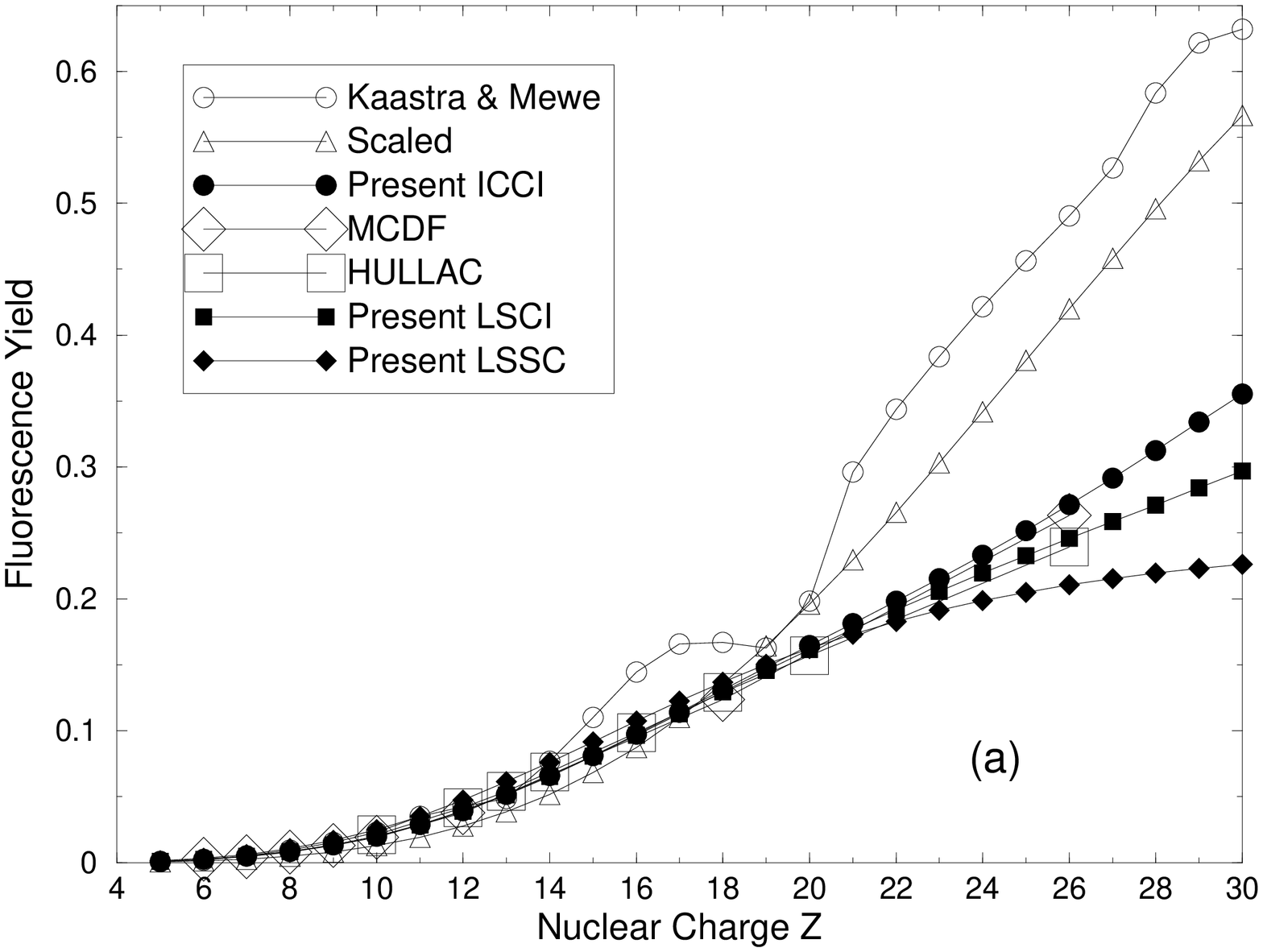,height=3.in,angle=-90.}}
\caption{(a) Comparison of various computed and inferred fluorescence yields $\xi$
for Be-like $1s2s^22p$ ions:
present LS results in the single-configuration (SC) approximation -- solid diamonds; present LS results with configuration interaction (CI) included -- solid squares; 
present intermediate coupling (IC) results with CI included -- solid circles; HULLAC results from \cite{behar} -- open squares;
MCDF results from \cite{chen} -- open diamonds; \cite{k&m} -- open circles;
results when we scale \cite{mcguire69} \ion{B}{2} results, using Eq.~\ref{eqfy}
and the $\omega$ from \cite{lotz67,lotz68} -- open triangles. 
(b) Same as (a) focusing on the low-$Z$ region.
}
\centerline{\psfig{file=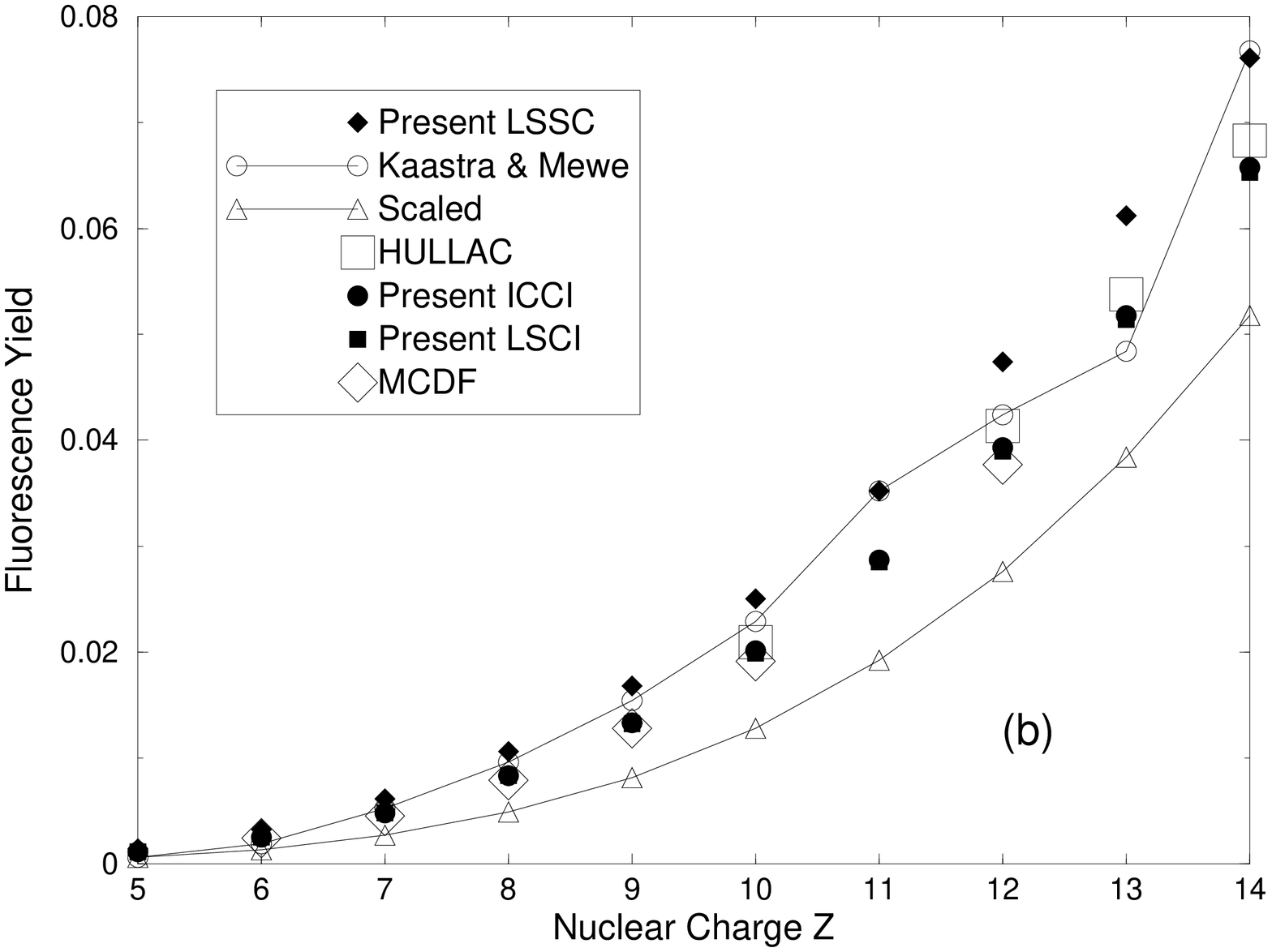,height=3.in,angle=-90.}}
\end{figure*}

\begin{figure*}
\centerline{\psfig{file=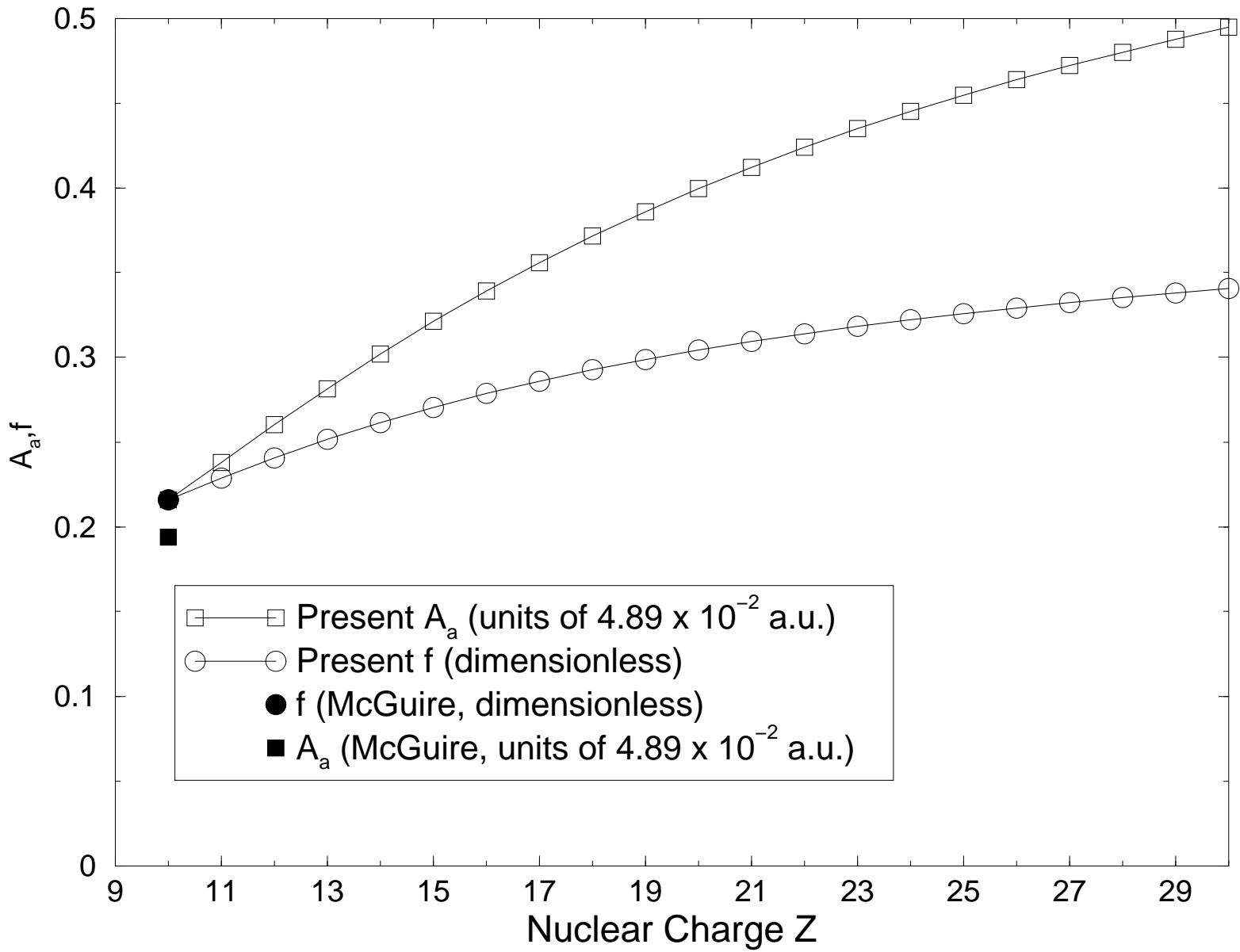,angle=-90.}}
\caption{Present LS emission oscillator strengths $f$ (dimensionless, open circles) and autoionization rates
$A_a$ (in units of $4.89\times 10^{-2}$ a.u., open squares) 
for F-like $1s2s^22p^6(^2S)$ ions
as a function
of the nuclear charge $Z$.  The emission oscillator strength and autoionization rate
from \cite{mcguire69} for \ion{Ne}{2},
as used by \cite{k&m}, are shown by the solid circle and square,
respectively.}
\end{figure*}

\begin{figure*}
\centerline{\psfig{file=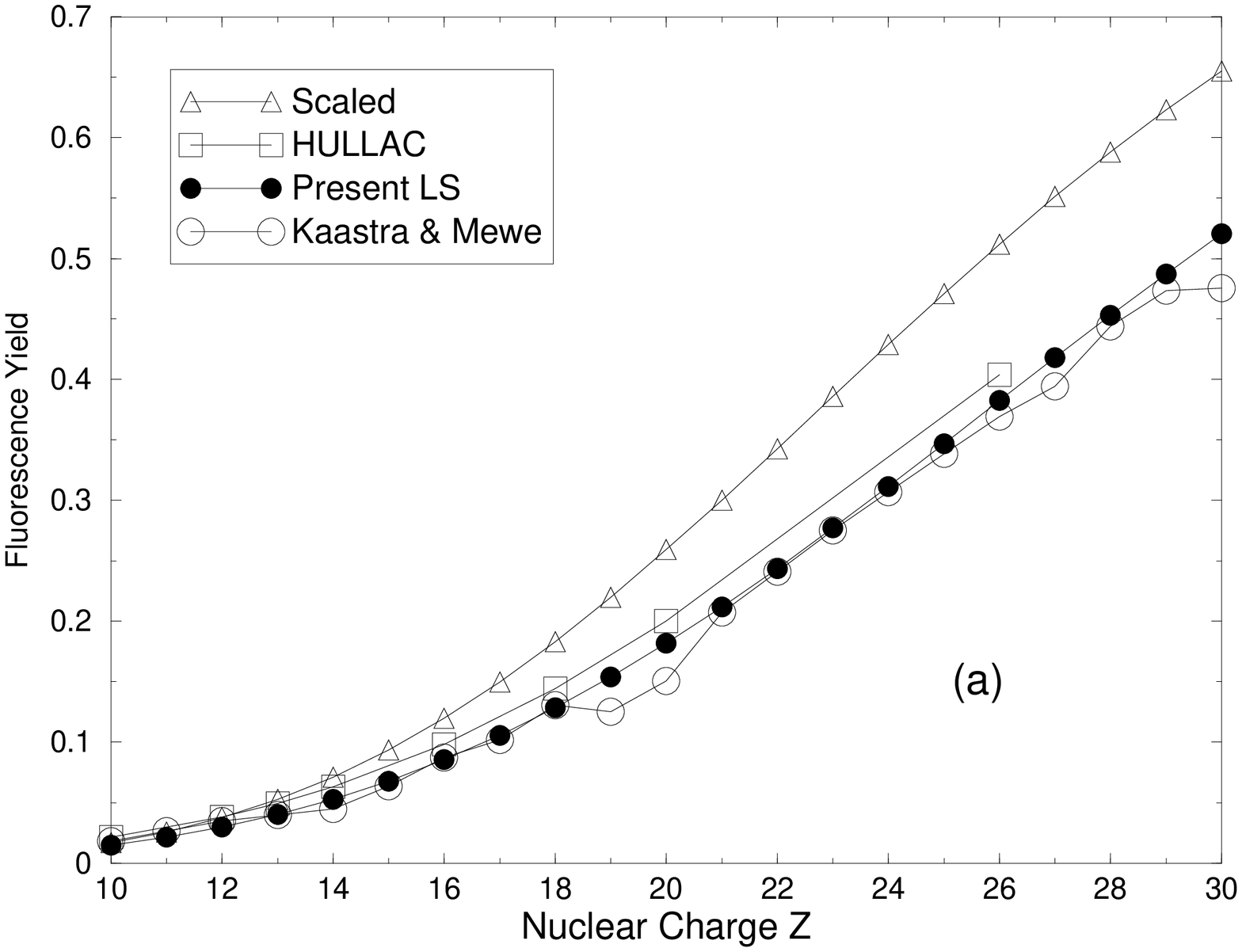,height=4.in,angle=-90.}}
\caption{(a) Comparison of various computed and inferred fluorescence yields $\xi$
for F-like $1s2s^22p^6(^2S)$ ions:
present LS results -- solid circles;  \cite{k&m} -- open circles;
results when we scale \cite{mcguire69} \ion{Ne}{2} results using Eq.~\ref{eqfy}
and the $\omega$ from \cite{lotz67,lotz68} -- open triangles;
HULLAC results of \citet{behar} -- open squares.
(b) Same as (a) focusing on the low-$Z$ region.
}
\centerline{\psfig{file=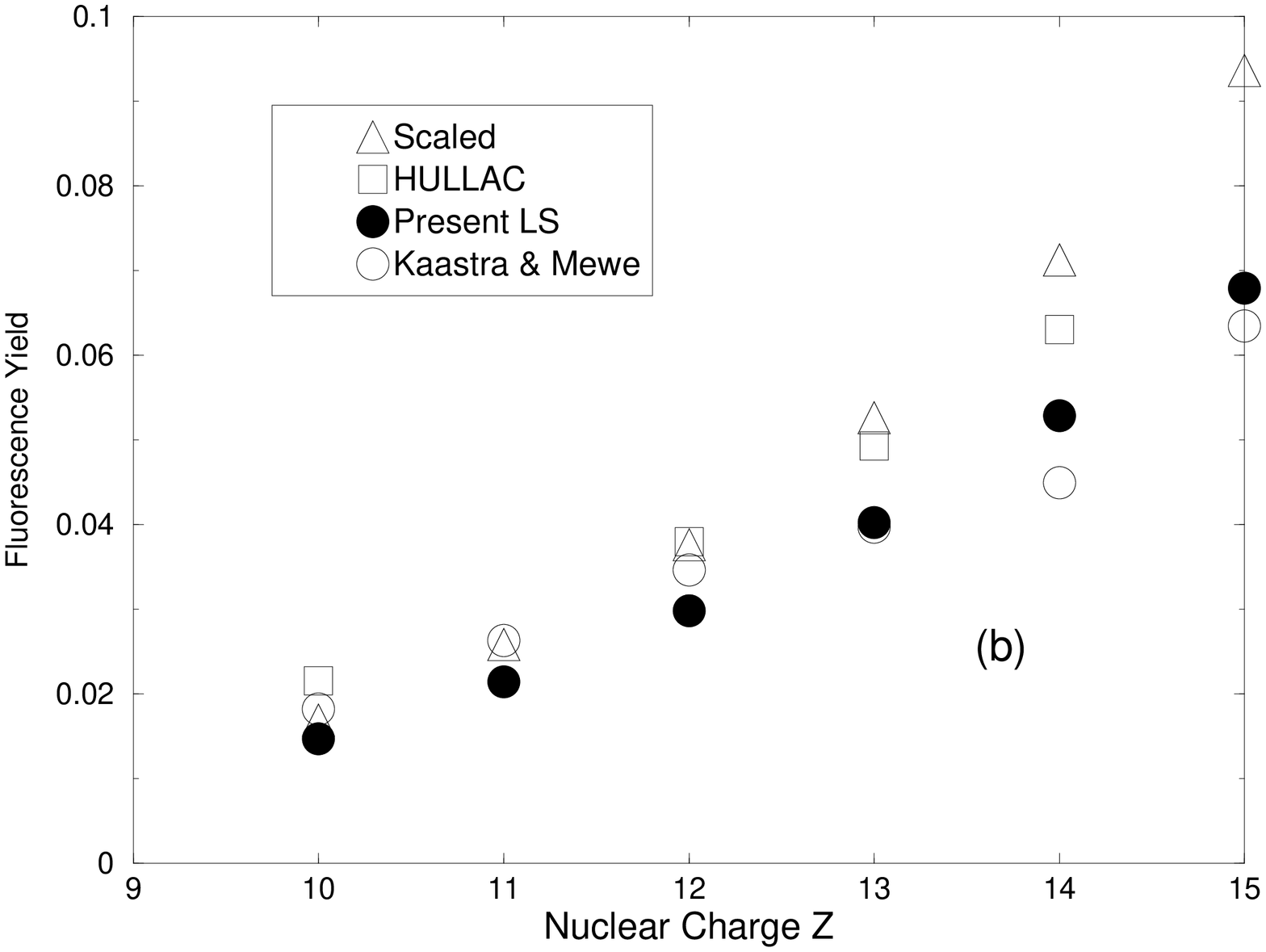,height=4.in,angle=-90.}}
\end{figure*}

\end{document}